\documentclass[letterpaper, 10 pt, conference]{ieeeconf}  

\IEEEoverridecommandlockouts                              

\usepackage{graphics} 
\usepackage{amsfonts}
\usepackage{import}
\usepackage{multirow,array,graphicx}
\usepackage{tabularx,booktabs}
\usepackage[table]{xcolor}

\newcommand{\refeq}[1]{(\ref{#1})}

\usepackage{physics}
\newcommand{\matr}[1]{\boldsymbol{#1}}
\newcommand{\vek}[1]{\boldsymbol{#1}}

\DeclareSymbolFont{matha}{OML}{txmi}{m}{it}
\DeclareMathSymbol{\varv}{\mathord}{matha}{118}

\newcommand{\pdrev}[2]{\dfrac{\partial #1}{\partial #2}}

\newcommand{\Rey}{\mathrm{\operatorname{\mathrm{R\kern-.04em e}}} }
\newcommand{\Pecl}{\mathrm{\operatorname{\mathrm{P\kern-.08em e}}}}
\newcommand{\Pran}{\mathrm{\operatorname{\mathrm{P\kern-.03em r}}}}
\newcommand{\Rayl}{\mathrm{\operatorname{\mathrm{R\kern-.04em a}}}}
\newcommand{\Nuss}{\mathrm{\operatorname{\mathrm{N\kern-.09em u}}}}
\newcommand{\Gras}{\mathrm{\operatorname{\mathrm{G\kern-.05em r}}}}
\newcommand{\Shear}{\mathrm{\operatorname{\mathrm{S\kern-.05em h}}}}
\newcommand{\Schm}{\mathrm{\operatorname{\mathrm{S\kern-.05em c}}}}

\newcommand{\Svec}{\vek{q}}
\newcommand{\Uvec}{\vek{u}}

\newcommand{\udim}{5}

\newcommand{\np}{N}
\newcommand{\nr}{n}

\newcommand{\Ar}{ \vek{\hat{A}} }
\newcommand{\Qr}{  \vek{\mathcal{\hat{Q}}}^{[1]} }
\newcommand{\Qrf}{ \vek{T} \vek{\mathcal{Q}}^{[1]} (\vek{I}_{m\times m} \otimes \vek{V}) }

\newcommand{\Br}{ \vek{\hat{B}} }
\newcommand{\Cr}{ \vek{\hat{C}} }

\newcommand{\Qf}{ \vek{\mathcal{Q}}^{[1]} }

\usepackage{siunitx}
\newcommand{\UnitBrackets}[1]{\,[ \si{#1} ]}

\newcommand{\peclet}{P\'{e}clet\ }

\newcommand{\refappendix}[1]{\hyperref[#1]{Appendix~\ref*{#1}}}

\usepackage{pgf}
\definecolor{lime}{HTML}{A6CE39}

\definecolor{lime}{HTML}{A6CE39}

\usepackage{pgf,tikz,pgfplots}
\pgfplotsset{compat=newest}
\usetikzlibrary{arrows,decorations.markings}
\usetikzlibrary{positioning}
\usepackage{tikz-network}
\usetikzlibrary{plotmarks}

\definecolor{mycolor1}{rgb}{0.25098,0.49804,0.71765}%

\usepackage[customcolors]{hf-tikz}
\usetikzlibrary{fit}
\tikzset{style green/.style={
		set fill color=green!20!lime!60,
		set border color=white,
	},
	style cyan/.style={
		set fill color=cyan!90!blue!60,
		set border color=white,
	},
	style orange/.style={
		set fill color=orange!80!red!60,
		set border color=white,
	},
	hor/.style={
		above left offset={-0.15,0.31},
		below right offset={0.15,-0.125},
		#1
	},
	ver/.style={
		above left offset={-0.1,0.3},
		below right offset={0.15,-0.15},
		#1
	}
}

\tikzset{%
	highlight/.style={rectangle,rounded corners,fill=red!80,fill opacity=0.3,thin,inner sep=0pt}
}

\graphicspath{{Figures/}}

\usepackage{pgf,tikz,pgfplots}
\pgfplotsset{compat=newest}
\usetikzlibrary{arrows,decorations.markings}
\usetikzlibrary{positioning}
\usepackage{tikz-network}
\usetikzlibrary{plotmarks}

\title{\LARGE \bf
Model Order Reduction of The Time-Dependent Advection-Diffusion-Reaction Equation with Time-Varying Coefficients: Application to Real-Time Water Quality Monitoring
}

\author{Ahmed Elkhashap$^{1}$ and Dirk Abel$^{1}$
\thanks{$^{1}$Ahmed Elkhashap and Dirk Abel are with Institute of Automatic Control, RWTH Aachen University, 52062 Aachen, Germany {\tt\small a.elkhashap@irt.rwth-aachen.de}}%
}

\newcommand\copyrighttext{%
  
  \textcopyright 2022 IEEE. Personal use of this material is permitted.
  Permission from IEEE must be obtained for all other uses, in any current or future
  media, including reprinting/republishing this material for advertising or promotional
  purposes, creating new collective works, for resale or redistribution to servers or
  lists, or reuse of any copyrighted component of this work in other works.}
\newcommand\copyrightnotice{%
\begin{tikzpicture}[remember picture,overlay]
\node[anchor=north,yshift=-10pt] at (current page.north) {\shortstack{Accepted for publication at 2022 European Control Conference\\ \\
		\fbox{\parbox{\dimexpr\textwidth-\fboxsep-\fboxrule\relax}{\copyrighttext}}} };
\end{tikzpicture}%
}

\begin{document}
\copyrightnotice
\maketitle
\thispagestyle{empty}
\pagestyle{empty}

\begin{abstract}

Advection-Diffusion-Reaction (ADR) Partial Differential Equations (PDEs) appear in a wide spectrum of applications such as chemical reactors, concentration flows, and biological systems. A large number of these applications require the solution of ADR equations involving time-varying coefficients, where analytical solutions are usually intractable. Numerical solutions on the other hand require fine discretization and are computationally very demanding. Consequently, the models are normally not suitable for real-time monitoring and control purposes. In this contribution, a reduced order modeling method for a general ADR system with time-varying coefficients is proposed. Optimality of the reduced order model regarding the reduction induced error is achieved by using an $\mathcal{H}_2$-norm reduction method. The efficacy of the method is demonstrated using two test cases. Namely, a case for an ADR with arbitrary dynamics varying coefficients and a second case including the modeling of an exemplary water quality distribution path with randomly generated demand. The reduced order models are evaluated against high fidelity simulations using MATLAB's finite element method PDE toolbox. It is shown that the reduction can achieve a significant computational speedup allowing for the usage of the model for real-time applications with sampling times in milliseconds range. Moreover, the constructed ROM is shown to achieve high prediction accuracy with the normalized mean square error below $2.3\,\%$ for a real-world water quality simulation test case.
\end{abstract}

\section{INTRODUCTION}
Advection Diffusion Reaction (ADR) Partial Differential Equations (PDEs) are widely used in a lot of disciplines for modeling of various phenomena. The PDEs found in these various application is however deviant from the standardized linear-constant-parameter versions for which analytical solutions exist. Consequently, numerical solution is commonly employed which is usually computationally very expensive and inadequate to real-time applications. Nevertheless, solving ADR PDEs in real time is critical in achieving monitoring and control goals for numerous applications \cite{Elkhashap.2019b,Elkhashap.2021b}. For example, one of these applications is water quality monitoring in water distribution networks which gained a lot of momentum in the past few decades. Open source as well as commercial tools propose several methods for water quality modeling all while considering the used models computational complexity \cite{WQ_Epanet,rossman1999epanet,RTP_Epanet_1}. However, such models are mostly constrained by the highly simplifying assumptions and can usually only handle a restricted set of operating conditions. For example until now the widely used EPANET tool developed by the U.S. Environmental Protection Agency \cite{rossman1999epanet} for hydraulic and water quality simulation of water distribution network uses a simplified ADR equation neglecting the diffusion phenomena in disinfectant propagation. Moreover, the classical junction-link \cite{rossman1999epanet} representation of the water network disregards an accurate representation of the spatial variability of the properties under consideration. Other methods exist, which consider various effects such as diffusion, e.g. \cite{SHang_ADR}, however such methods normally suffer from high computational complexity specially when a fine spatial discretization is adopted. Tools offering high spatial resolution solutions combined with real-time suitable computational effort are to the authors knowledge still not well investigated. In most of the present tools the focus is either laid upon accurate solutions considering the spatial variability, e.g. Computational Fluid Dynamics (CFD) and Finite Element Method (FEM) \cite{ABOKIFA2016107,matlabPDEtool} tools, or the real time capability of the solution method through crude approximation, e.g. as in crude spatial lumping in EPANET \cite{rossman1999epanet}. Recent efforts for the consideration of the water quality with chosen spatial resolution for real-time control purposes can be also found. For example in \cite{WQModels4MPC} state space models of water networks are constructed with direct spatio-temporal discretization of ADR PDEs and then directly used for model-based control. Moreover, efforts regarding the computational complexity reduction can be found. For example in \cite{WQMOR} the network state-space models are joined and reduced using adapted projection based techniques. However, the strategy of collectively modelling water networks with direct spatio-temporal discretization is unsuitable for a network size scale-up or fine discretization cases. Moreover, the employed Model Order Reduction (MOR) technique for such Network monolithic models causes the inability to fully preserve the structure, properties and parametric dependence of the Full Order Model (FOM) within the Reduced Order Model (ROM). Hence special adaptations in the reduction method are needed to guarantee the conservation of the essential properties, e.g. the ROM stability in \cite{WQMOR}. In this contribution, a general approach for the solution and model order reduction of ADR-PDEs with time-varying coefficients is proposed. The approach is based on using augmentations in the FOM formulation reaching a standard bilinear form on which system theoretic MOR methods for bilinear systems can be employed \cite{MORbilinH2,Elkhashap.2021a,Elkhashap2022realtime}. The proposed approach preserves the FOM structural interpretability and dependency on the time varying coefficients allowing for its use in real-time estimation and control purposes. Moreover, the proposed method inherits by construction the qualities of minimal reduction error, stability, and error bounds from the employed $\mathcal{H}_2$ norm MOR for bilinear systems \cite{redmann2021bilinear}. Furthermore, the approach is evaluated empirically in two examples including one for water quality modeling comparing the ROM solution to a FOM solution using the MATLAB's PDE FEM solver \cite{matlabPDEtool} as ground truth. In the water quality modeling example the EPANET tool solution is used in comparison highlighting the effect of the high spatial resolution and inclusion of the diffusion effects. The contribution is organized as following. First the general ADR PDE with time varying coefficients is introduced in section \ref{sec:1} highlighting the adopted solution method reaching a standard bilinear form. Second in section \ref{sec:2}, the MOR method is briefly introduced illustrating the produced ROM. Third in section \ref{sec:3}, the two evaluation scenarios of the ROM is are illustrated also introducing the water quality model utilized for the second scenario. Finally, the results are shown and discussed in section \ref{sec:4} followed by a brief conclusion.
\section{Methods}
\subsection{ADR PDE with time varying coefficients}\label{sec:1}
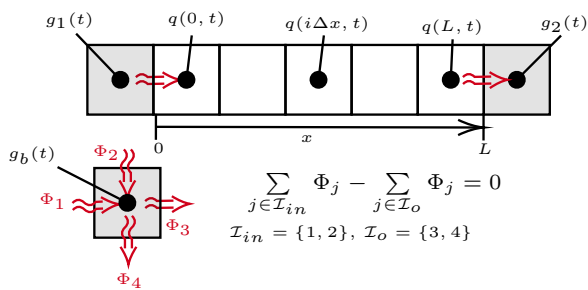
\begin{figure}[b!]
	\centering
	\vspace{-0.5cm}
	\resizebox{\linewidth}{!}{
	\tikzset{every picture/.style={line width=0.75pt}} 

\begin{tikzpicture}[x=0.75pt,y=0.75pt,yscale=-1,xscale=1]

\draw    (193.83,58.67) -- (334.8,58.96) ;
\draw [shift={(334.8,58.96)}, rotate = 180.12] [color={rgb, 255:red, 0; green, 0; blue, 0 }  ][line width=0.75]    (0,5.59) -- (0,-5.59)(10.93,-3.29) .. controls (6.95,-1.4) and (3.31,-0.3) .. (0,0) .. controls (3.31,0.3) and (6.95,1.4) .. (10.93,3.29)   ;
\draw [shift={(193.83,58.67)}, rotate = 180.12] [color={rgb, 255:red, 0; green, 0; blue, 0 }  ][line width=0.75]    (0,5.59) -- (0,-5.59)   ;
\draw [color={rgb, 255:red, 208; green, 2; blue, 27 }  ,draw opacity=1 ]   (185.22,37.31) .. controls (186.86,35.62) and (188.53,35.59) .. (190.22,37.22) -- (194.37,37.15) -- (197.37,37.09)(185.28,40.31) .. controls (186.91,38.62) and (188.58,38.59) .. (190.28,40.22) -- (194.43,40.15) -- (197.43,40.09) ;
\draw [shift={(204.4,38.47)}, rotate = 538.97] [color={rgb, 255:red, 208; green, 2; blue, 27 }  ,draw opacity=1 ][line width=0.75]    (7.65,-3.43) .. controls (4.86,-1.61) and (2.31,-0.47) .. (0,0) .. controls (2.31,0.47) and (4.86,1.61) .. (7.65,3.43)   ;
\draw    (209.26,19.3) -- (206.96,38.41) ;
\draw [shift={(206.96,38.41)}, rotate = 96.89] [color={rgb, 255:red, 0; green, 0; blue, 0 }  ][fill={rgb, 255:red, 0; green, 0; blue, 0 }  ][line width=0.75]      (0, 0) circle [x radius= 3.35, y radius= 3.35]   ;
\draw  [fill={rgb, 255:red, 0; green, 0; blue, 0 }  ,fill opacity=0.11 ] (164.33,23.41) -- (192.75,23.41) -- (192.75,53.4) -- (164.33,53.4) -- cycle ;
\draw   (192.75,23.41) -- (221.17,23.41) -- (221.17,53.4) -- (192.75,53.4) -- cycle ;
\draw   (221.17,23.41) -- (249.58,23.41) -- (249.58,53.4) -- (221.17,53.4) -- cycle ;
\draw   (306.49,23.41) -- (334.91,23.41) -- (334.91,53.4) -- (306.49,53.4) -- cycle ;
\draw   (278.07,23.41) -- (306.49,23.41) -- (306.49,53.4) -- (278.07,53.4) -- cycle ;
\draw  [fill={rgb, 255:red, 0; green, 0; blue, 0 }  ,fill opacity=0.11 ] (334.91,23.41) -- (363.32,23.41) -- (363.32,53.4) -- (334.91,53.4) -- cycle ;
\draw    (321.5,20.98) -- (320.7,38.41) ;
\draw [shift={(320.7,38.41)}, rotate = 92.62] [color={rgb, 255:red, 0; green, 0; blue, 0 }  ][fill={rgb, 255:red, 0; green, 0; blue, 0 }  ][line width=0.75]      (0, 0) circle [x radius= 3.35, y radius= 3.35]   ;
\draw    (159.5,18.55) -- (178.54,38.41) ;
\draw [shift={(178.54,38.41)}, rotate = 46.21] [color={rgb, 255:red, 0; green, 0; blue, 0 }  ][fill={rgb, 255:red, 0; green, 0; blue, 0 }  ][line width=0.75]      (0, 0) circle [x radius= 3.35, y radius= 3.35]   ;
\draw    (368.16,19.36) -- (349.11,38.41) ;
\draw [shift={(349.11,38.41)}, rotate = 135] [color={rgb, 255:red, 0; green, 0; blue, 0 }  ][fill={rgb, 255:red, 0; green, 0; blue, 0 }  ][line width=0.75]      (0, 0) circle [x radius= 3.35, y radius= 3.35]   ;
\draw   (249.58,23.41) -- (278,23.41) -- (278,53.4) -- (249.58,53.4) -- cycle ;
\draw    (266.1,19.3) -- (263.79,38.41) ;
\draw [shift={(263.79,38.41)}, rotate = 96.89] [color={rgb, 255:red, 0; green, 0; blue, 0 }  ][fill={rgb, 255:red, 0; green, 0; blue, 0 }  ][line width=0.75]      (0, 0) circle [x radius= 3.35, y radius= 3.35]   ;
\draw [color={rgb, 255:red, 208; green, 2; blue, 27 }  ,draw opacity=1 ]   (189.22,90.68) .. controls (190.85,88.99) and (192.52,88.96) .. (194.22,90.59) -- (198.37,90.52) -- (201.37,90.46)(189.27,93.68) .. controls (190.91,91.99) and (192.58,91.96) .. (194.27,93.59) -- (198.42,93.51) -- (201.42,93.46) ;
\draw [shift={(208.4,91.83)}, rotate = 538.97] [color={rgb, 255:red, 208; green, 2; blue, 27 }  ,draw opacity=1 ][line width=0.75]    (7.65,-3.43) .. controls (4.86,-1.61) and (2.31,-0.47) .. (0,0) .. controls (2.31,0.47) and (4.86,1.61) .. (7.65,3.43)   ;
\draw [color={rgb, 255:red, 208; green, 2; blue, 27 }  ,draw opacity=1 ]   (325.94,37.25) .. controls (327.57,35.56) and (329.24,35.53) .. (330.94,37.16) -- (335.09,37.09) -- (338.09,37.03)(325.99,40.25) .. controls (327.62,38.56) and (329.29,38.53) .. (330.99,40.16) -- (335.14,40.09) -- (338.14,40.03) ;
\draw [shift={(345.11,38.41)}, rotate = 538.97] [color={rgb, 255:red, 208; green, 2; blue, 27 }  ,draw opacity=1 ][line width=0.75]    (7.65,-3.43) .. controls (4.86,-1.61) and (2.31,-0.47) .. (0,0) .. controls (2.31,0.47) and (4.86,1.61) .. (7.65,3.43)   ;
\draw  [fill={rgb, 255:red, 0; green, 0; blue, 0 }  ,fill opacity=0.11 ] (167.33,76.41) -- (195.75,76.41) -- (195.75,106.4) -- (167.33,106.4) -- cycle ;
\draw [color={rgb, 255:red, 208; green, 2; blue, 27 }  ,draw opacity=1 ]   (183.5,97) .. controls (185.17,98.67) and (185.17,100.33) .. (183.5,102) .. controls (181.83,103.67) and (181.83,105.33) .. (183.5,107) -- (183.5,108) -- (183.5,111)(180.5,97) .. controls (182.17,98.67) and (182.17,100.33) .. (180.5,102) .. controls (178.83,103.67) and (178.83,105.33) .. (180.5,107) -- (180.5,108) -- (180.5,111) ;
\draw [shift={(182,118)}, rotate = 270] [color={rgb, 255:red, 208; green, 2; blue, 27 }  ,draw opacity=1 ][line width=0.75]    (7.65,-3.43) .. controls (4.86,-1.61) and (2.31,-0.47) .. (0,0) .. controls (2.31,0.47) and (4.86,1.61) .. (7.65,3.43)   ;
\draw [color={rgb, 255:red, 208; green, 2; blue, 27 }  ,draw opacity=1 ]   (157.93,91.5) .. controls (159.51,89.75) and (161.17,89.67) .. (162.92,91.25) .. controls (164.67,92.83) and (166.33,92.75) .. (167.91,91) -- (167.94,91) -- (170.93,90.85)(158.07,94.5) .. controls (159.66,92.75) and (161.32,92.67) .. (163.07,94.25) .. controls (164.82,95.83) and (166.48,95.75) .. (168.06,94) -- (168.09,94) -- (171.08,93.85) ;
\draw [shift={(178,92)}, rotate = 537.14] [color={rgb, 255:red, 208; green, 2; blue, 27 }  ,draw opacity=1 ][line width=0.75]    (7.65,-3.43) .. controls (4.86,-1.61) and (2.31,-0.47) .. (0,0) .. controls (2.31,0.47) and (4.86,1.61) .. (7.65,3.43)   ;
\draw [color={rgb, 255:red, 208; green, 2; blue, 27 }  ,draw opacity=1 ]   (183.04,68.41) .. controls (184.71,70.08) and (184.71,71.74) .. (183.04,73.41) .. controls (181.37,75.08) and (181.37,76.74) .. (183.04,78.41) -- (183.04,79.41) -- (183.04,82.41)(180.04,68.41) .. controls (181.71,70.08) and (181.71,71.74) .. (180.04,73.41) .. controls (178.37,75.08) and (178.37,76.74) .. (180.04,78.41) -- (180.04,79.41) -- (180.04,82.41) ;
\draw [shift={(181.54,89.41)}, rotate = 270] [color={rgb, 255:red, 208; green, 2; blue, 27 }  ,draw opacity=1 ][line width=0.75]    (7.65,-3.43) .. controls (4.86,-1.61) and (2.31,-0.47) .. (0,0) .. controls (2.31,0.47) and (4.86,1.61) .. (7.65,3.43)   ;

\draw (268.6,14.41) node  [font=\tiny]  {$q( i\Delta x,t) \ $};
\draw (212.46,12.3) node  [font=\tiny]  {$q( 0,t) \ $};
\draw (335.41,68) node  [font=\tiny]  {$L$};
\draw (258.89,63.95) node  [font=\tiny]  {$x$};
\draw (322.42,15.54) node  [font=\tiny]  {$q( L,t) \ $};
\draw (157.75,12.71) node  [font=\tiny]  {$g_{1}( t) \ $};
\draw (370.7,14.33) node  [font=\tiny]  {$g_{2}( t) \ $};
\draw (195,68) node  [font=\tiny]  {$0$};
\draw (288,88) node  [font=\scriptsize]  {$\sum\limits_{j\in \mathcal{I}_{in}} \Phi_{j} -\sum\limits_{j\in \mathcal{I}_o} \Phi_{j}=0$};
\draw (183.75,125) node  [font=\tiny,color={rgb, 255:red, 208; green, 2; blue, 27 }  ,opacity=1 ]  {$\Phi _{4} \ $};
\draw (150.75,90.82) node  [font=\tiny,color={rgb, 255:red, 208; green, 2; blue, 27 }  ,opacity=1 ]  {$\Phi _{1} \ $};
\draw (173.75,69.82) node  [font=\tiny,color={rgb, 255:red, 208; green, 2; blue, 27 }  ,opacity=1 ]  {$\Phi _{2} \ $};
\draw (204.75,101.82) node  [font=\tiny,color={rgb, 255:red, 208; green, 2; blue, 27 }  ,opacity=1 ]  {$\Phi _{3} \ $};
\draw (141.75,70.82) node  [font=\tiny]  {$g_{b}( t) \ $};
\draw (277.75,104.82) node  [font=\tiny]  {$\mathcal{I}_{in}=\{1,2\},\, \mathcal{I}_o=\{3,4\}$};
\draw    (151.5,76) -- (181.54,91.41) ;
\draw [shift={(181.54,91.41)}, rotate = 27.15] [color={rgb, 255:red, 0; green, 0; blue, 0 }  ][fill={rgb, 255:red, 0; green, 0; blue, 0 }  ][line width=0.75]      (0, 0) circle [x radius= 3.35, y radius= 3.35]   ;

\end{tikzpicture}
	}
	\vspace{-0.8cm}
	\caption{Schematic of ADR equation employed spatial discretization and boundary conditions handling}
	\vspace{-0.5cm}
	\label{fig:ADR_Disk}
\end{figure}
The main equation under consideration is the inhomogeneous ADR equation in one spatial dimension defined over a finite domain $x\in [0,L]$. The PDE governing a property $q(x,t),\quad t \in \mathbb{R}^+$ can be expressed as follows
\begin{equation}
    \pdrev{q}{t}=-v(t)\pdrev{q}{x}+D(t)\pdrev{^2 q}{x^2}+r(t)q+s(t),\label{eq:ADR_PDE}
\end{equation}
with the propagation velocity coefficient $v(t)$, Diffusion coefficient $D(t)$, reaction rate $r(t)$, and source term $s(t)$. The property values at the boundaries are assumed to be directly known (Dirichlet boundary condition). This is as the junctions of the PDE domain (boundaries) are treated separately as illustrated in Fig. \ref{fig:ADR_Disk} using a mass conserving Robin-type boundary condition (cf. open boundary in \cite{Danckwerts.1995})
\begin{align}
\sum\limits_{j\in \mathcal{I}_{in}} \Phi_{j}(x_b,t) -\sum\limits_{j\in \mathcal{I}_o} \Phi_{j}( x_b,t)=0,\\
\Phi(x,t)=v(t)q(x,t)-D(t)\pdrev{q}{x}(x,t)\label{eq:FluxFcn},
\end{align}
with the flux $\Phi(x,t)$ (\ref{eq:FluxFcn}) crossing the respective boundary at $x_b$, the set of the property inflow $\mathcal{I}_{in}$, and set of the property outflow $\mathcal{I}_{o}$. The partial derivatives of the property in the flux functions are approximated using finite differences after introducing the function for the property at the respective boundary $g_b(t)$. Hence, for junctions with multiple in- and outflux algebraic conservation equations are to be solved (as shown in Fig. \ref{fig:ADR_Disk}). The boundary conditions are defined in this general way to allow constructing networks with arbitrary topology whose links are represented by ADR PDEs. The advantage of this formulation is that the diffusion across the junctions is considered, which is important for cases of networks with zero advection, e.g. dead ends in water distribution networks. Note that, the property state at the nodes of the network can also be represented with separate ODEs analogously if accumulation in junctions is to be considered. A finite difference based Method of Lines (semi-discretization) is applied for the PDE solution \cite{Abgrall.2017}. The spatial variable is discretized on a uniform grid of $\np$ points with segment length $\Delta x=\frac{L}{\np-1}$. The convective term is approximated using first order upwind scheme and the diffusive using central difference. The boundary conditions are handled using a ghost point technique. For illustrative simplicity, only the case of a single inflow flux $g(t)$ with positive flow velocity and an open right boundary is considered hereafter. Consequently, the following high dimensional nonlinear ODE system is produced
\begin{equation}
\dot{\vek{q}}=(v(t)\matr{Q}_1+D(t)\matr{Q}_2+r(t)\matr{Q}_3)\vek{q}+\matr{b}_1s(t) +\matr{b}_2{\varphi}(t),    
\end{equation}
with the vector $\vek{q} \in \mathbb{R}^{\np}:=[q(i\Delta x ,t)]^\mathrm{T},\quad i=\{1,\cdots,\np\}$
\begin{subequations}\label{eq:FOMBilin}
	\begin{equation}
	\matr{Q}_1=\frac{1}{\Delta x}\small{
		\begin{bmatrix} 
		-1   &0   &\ldots&0\\
		1   & -1  &\ddots&\vdots\\
		\vdots&\ddots&\ddots   &0\\
		0     &\ldots&1   &-1
		\end{bmatrix}}, 
	\matr{Q}_3=\matr{I},\,
	\vek{b}_1=\begin{bmatrix} 1\\	1\\	\vdots\\1\end{bmatrix},
	\end{equation}
	\begin{equation}
	\vek{Q_2}=\frac{1}{\Delta x^2}\small{
		\begin{bmatrix} 
		-2   &1   &\ldots&0\\
		1   & -2  &\ddots&\vdots\\
		\vdots&\ddots&\ddots   &1\\
		0     &\ldots&1   &-1
		\end{bmatrix}},\,
		\vek{b}_2=\frac{1}{\Delta x^2}\begin{bmatrix} 1 \\	0\\	\vdots\\0\end{bmatrix},
	\end{equation}
\end{subequations}
and the function $\varphi(v(t),g(t),D(t),q(\Delta x,t))$ representing the dependency due to the right boundary condition after discretization. This dependency is formulated according to the handling method, e.g. ghost point, Taylor-based expansion. A ghost point handling with partial derivative approximation at the right boundary point delivers $\varphi(t)=\Delta xv(t)g(t)+D(t)g(t)$. Now (\ref{eq:FOMBilin}) can be rearranged into a standard bilinear form by introducing the augmented input vector $\vek{u}:=[v(t),D(t),r(t)+1,s(t),\varphi(t)]^\mathrm{T}$
\begin{subequations}
\begin{equation}
    \dot{\Svec}=\matr{A}\Svec+\Qf\Uvec\otimes\Svec +\matr{B}\Uvec,\label{eq:FOM}
\end{equation}
\begin{equation}
\vek{y}=\matr{C}\Svec\label{eq:FOM_out}    
\end{equation}\label{eq:FOM_Tot}
\end{subequations}
with the system matrix $\matr{A} \in \mathbb{R}^{\np \times \np}=-\matr{I}$, input matrix $\matr{B}\in \mathbb{R}^{\np\times\udim}=[\matr{0}_{\np\times 3},\,\vek{b}_1,\vek{b}_2]$, and measurement matrix $\matr{C} \in \mathbb{R}^{\np \times\np}$ assumed to be identity $\matr{I}$ indicating that a measurement is available at each point. The bilinear term is expressed using the Kronecker product notation $\otimes$, where $\matr{\mathcal{Q}}^{[1]}\in \mathbb{R}^{\np\times 5\np}$ is the mode-1 matricization of the $3^{rd}$ order tensor $\matr{\mathcal{Q}}\in\mathbb{R}^{\np\times\np\times 5}$ having the matrices $\matr{Q}_i \in \mathbb{R}^{\np\times\np}, \forall i\in\{1,..,\udim\}$ as frontal slices, i.e. $\matr{\mathcal{Q}}^{[1]}=[\matr{Q}_1,\matr{Q}_2,\matr{Q}_3,\matr{0}_{\np\times3\np}]$. The choice of the augmented input vector allowed a system matrix $\matr{A}$ with strictly  negative  eigenvalues of $-1$ with algebraic multiplicity $N$. Hence, \refeq{eq:FOM_Tot} represents a Bounded Input Bounded Output (BIBO) stable bilinear system \cite{MORbilinH2Zhang.2002,Redmann2019TheML} in standard form. The next step is to perform a projection based MOR method with the lowest possible reduction error.

\subsection{Model Order Reduction}\label{sec:2}
\begin{figure}[h!]
  \centering
    \includegraphics[width=\linewidth]{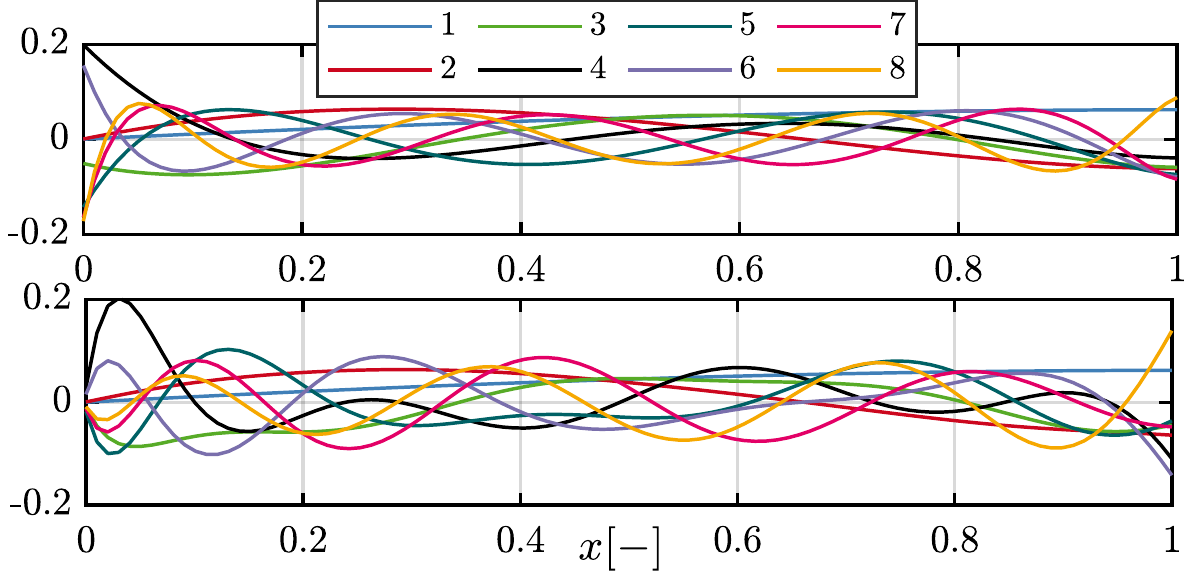}
    \vspace{-0.6cm}
  	\caption{Spatial Modes for the right (bottom) and left (top) projection matrices, i.e. columns of $\matr{V}$, rows of $\matr{T}$, resulting from the reduction method}
	\vspace{-0.3cm}
	\label{fig:MORModes}
\end{figure}

Classic projection-based reduction methods are mainly based on the time space separation assumption when considering the original PDE. Generally, A PDE solution satisfying the separation property can be expressed using an infinite series of the inner product between the spatial and temporal bases. Hence, the PDE solution can be approximated by an inner product between spatial and temporal modes. Namely, a truncation of such infinite series to only a finite number of terms with the selection of the most influential spatial and temporal modes delivers a good approximation of the solution. Empirical methods, e.g. POD-Galerkin, Trajectory Piece Wise Linear (TPWL) methods \cite{MORParamOverview}, rely on simulation data in finding the bases delivering the most accurate reconstruction of a full order dynamics. However, such methods (specially for nonlinear systems) have no guarantees nor bounds regarding the reduction error. Moreover, it is challenging in such methods to preserve the original system structure and parametric dependency. Hence these methods generally work when the ROM is used with the parameter and input profiles in the vicinity of the ones used in the reduction. In the context of water quality modeling and reduction special adaptations are needed to achieve a ROM inheriting the stability of the FOM \cite{WQMOR}. On the other hand, system theoretic methods introduce more rigor in the reduction procedure \cite{MORParamOverview}, where for some special cases notions of optimal reduction and ROM error bounds can be achieved. The main method of interest employed here is the $\mathcal{H}_{\text{2}}$ norm reduction method for bilinear systems. The main idea of the method is to apply a Petrov-Galerkin projection based reduction, i.e. double sided projection using the test and trial bases matrices $\matr{V},\, \matr{W} \in \mathbb{R}^{\np\times \nr}$, on the FOM delivering the following ROM equivalent
\begin{subequations}
\begin{equation}
\vek{{\dot{\hat q}}}=\Ar\vek{\hat q}+\Qr \vek{u}\otimes \vek{\hat q} + \Br \vek{u},\label{eq:ROM}
\end{equation}	
\begin{equation}
\vek{y}=\Cr \vek{\hat q}\label{eq:ROM_out}
\end{equation}
\end{subequations}
with the reduced state space vector $\vek{\hat q}\in \mathbb{R}^{\nr}$, reduced order system matrices $\Ar \in \mathbb{R}^{\nr\times \nr}$, $\Qr \in \mathbb{R}^{\nr \times \udim\nr}$, $\Br \in \mathbb{R}^{\nr \times \udim}$, and $\vek{\hat{C}} \in \mathbb{R}^{n\times \nr}$ as follows
\begin{align}
&\Ar=\vek{T}\vek{A}\vek{V},\quad \Br=\vek{T}\vek{B},\quad\Cr=\vek{C}\vek{V},\\
&\Qr=\Qrf,\quad \vek{T}=(\vek{W}^\mathrm{T}\vek{V})^{-1}\vek{W}^\mathrm{T}.
\end{align}
The error between the FOM and ROM is formulated in terms of the $\mathcal{H}_{\text{2}}$ norm then used to find the projection bases leading to minimum error. Several methods are proposed for finding the bases, e.g, as constrained minimization in \cite{MORbilinH2Zhang.2002}, iterative algorithms in \cite{MORbilinH2}. Here a slightly modified version of the Sylvester equation based algorithm in \cite{MORbilinH2} is used. The modification is mainly in the progression criteria of the algorithm iteration, where instead of using the change in the eigenvalues of the ROM matrix $\Ar$, the representation of the $\mathcal{H}_2$ norm of the error system proposed in \cite{Redmann2019TheML} (error bound expression) is calculated explicitly and used as the algorithm progression criteria. This adaptation introduces significant increase in the algorithm computational effort as the explicit error representation requires the explicit calculation of the FOM, ROM, and cross system Gramians. The algorithm computational efficiency is however in our case irrelevant as it is used only in an offline step. The main advantage of the method is the preservation of the FOM structure and parametric dependency without relying in any way on empirical simulation data. Moreover, recent results on the relation between the $\mathcal{H}_{\text{2}}$ norm error and the explicit output error including bounds on the output error can be found in \cite{redmann2021bilinear,Redmann2019TheML}. The reduction method is applied to a time and space scaled version of the FOM (\ref{eq:FOMBilin}) with $\np=500$ and $n=8$. The spatial modes emergent after the convergence of the reduction method are shown in Fig. \ref{fig:MORModes}.
Besides the theoretical guarantees of the ROM accuracy, the ROM is also empirically evaluated in a set of scenarios presented in the next section.
\section{Case Studies}\label{sec:3}
The method elaborated in the previous sections is now evaluated in two scenarios including one of real world applications namely:
\begin{enumerate}
    \item A general ADR equation with time-varying signals as its coefficients,
    \item A water quality modeling and simulation test case.
\end{enumerate}
The simulation of the FOM for the different scenarios is performed using Matlab's FEM based PDE toolbox \cite{matlabPDEtool}. The tool requires the PDE to be solved in divergence form. It also requires an explicit time functional form of the inputs time varying profiles. Hence, for the time varying input profiles either a functional approximation is used or interpolation routines are used for the piece-wise changing functions, e.g. demand patterns from the EPANET software. For the implementation of the ROM time integration schemes CasADi \cite{Andersson2019} toolbox is employed utilizing the 3rd order legendre collocation scheme. In the following the two scenarios are elaborated and then the results are presented.  

\subsection{Simulation-based Evaluations}
For the first scenario the ROM of the general ADR model is evaluated against a high fidelity simulation for a wide range of flow conditions with varying \peclet number. Moreover, a classical POD-Galerkin reduction (being the standard method for nonlinear systems) is compared against the proposed approach highlighting the efficacy of the proposed method. The time-varying coefficients are chosen  to be either chirp signals of varying amplitude and frequency or a superposition of harmonics, steps, exponential growths, and decays. 
\subsection{Water Quality Modeling Example}
\begin{figure}[h!]
	\centering
	\resizebox{\linewidth}{!}{
	\tikzset{every picture/.style={line width=0.75pt}} 

\begin{tikzpicture}[x=0.75pt,y=0.75pt,yscale=-1,xscale=1]

\draw    (171.83,25.67) -- (452.5,25.67) ;
\draw    (207.18,25.67) ;
\draw [shift={(207.18,25.67)}, rotate = 0] [color={rgb, 255:red, 0; green, 0; blue, 0 }  ][fill={rgb, 255:red, 0; green, 0; blue, 0 }  ][line width=0.75]      (0, 0) circle [x radius= 3.35, y radius= 3.35]   ;
\draw    (234.44,25.67) ;
\draw [shift={(234.44,25.67)}, rotate = 0] [color={rgb, 255:red, 0; green, 0; blue, 0 }  ][fill={rgb, 255:red, 0; green, 0; blue, 0 }  ][line width=0.75]      (0, 0) circle [x radius= 3.35, y radius= 3.35]   ;
\draw    (316.22,25.67) ;
\draw [shift={(316.22,25.67)}, rotate = 0] [color={rgb, 255:red, 0; green, 0; blue, 0 }  ][fill={rgb, 255:red, 0; green, 0; blue, 0 }  ][line width=0.75]      (0, 0) circle [x radius= 3.35, y radius= 3.35]   ;
\draw    (288.96,25.67) ;
\draw [shift={(288.96,25.67)}, rotate = 0] [color={rgb, 255:red, 0; green, 0; blue, 0 }  ][fill={rgb, 255:red, 0; green, 0; blue, 0 }  ][line width=0.75]      (0, 0) circle [x radius= 3.35, y radius= 3.35]   ;
\draw    (261.7,25.67) ;
\draw [shift={(261.7,25.67)}, rotate = 0] [color={rgb, 255:red, 0; green, 0; blue, 0 }  ][fill={rgb, 255:red, 0; green, 0; blue, 0 }  ][line width=0.75]      (0, 0) circle [x radius= 3.35, y radius= 3.35]   ;
\draw    (343.48,25.67) ;
\draw [shift={(343.48,25.67)}, rotate = 0] [color={rgb, 255:red, 0; green, 0; blue, 0 }  ][fill={rgb, 255:red, 0; green, 0; blue, 0 }  ][line width=0.75]      (0, 0) circle [x radius= 3.35, y radius= 3.35]   ;
\draw    (370.74,25.67) ;
\draw [shift={(370.74,25.67)}, rotate = 0] [color={rgb, 255:red, 0; green, 0; blue, 0 }  ][fill={rgb, 255:red, 0; green, 0; blue, 0 }  ][line width=0.75]      (0, 0) circle [x radius= 3.35, y radius= 3.35]   ;
\draw    (452.5,25.67) ;
\draw [shift={(452.5,25.67)}, rotate = 0] [color={rgb, 255:red, 0; green, 0; blue, 0 }  ][fill={rgb, 255:red, 0; green, 0; blue, 0 }  ][line width=0.75]      (0, 0) circle [x radius= 3.35, y radius= 3.35]   ;
\draw    (425.26,25.67) ;
\draw [shift={(425.26,25.67)}, rotate = 0] [color={rgb, 255:red, 0; green, 0; blue, 0 }  ][fill={rgb, 255:red, 0; green, 0; blue, 0 }  ][line width=0.75]      (0, 0) circle [x radius= 3.35, y radius= 3.35]   ;
\draw    (398,25.67) ;
\draw [shift={(398,25.67)}, rotate = 0] [color={rgb, 255:red, 0; green, 0; blue, 0 }  ][fill={rgb, 255:red, 0; green, 0; blue, 0 }  ][line width=0.75]      (0, 0) circle [x radius= 3.35, y radius= 3.35]   ;
\draw  [fill={rgb, 255:red, 184; green, 233; blue, 134 }  ,fill opacity=1 ] (166.4,19.35) -- (177.27,19.35) -- (177.27,31.99) -- (166.4,31.99) -- cycle ;
\draw  [fill={rgb, 255:red, 184; green, 233; blue, 134 }  ,fill opacity=1 ] (407.23,50.35) -- (418.1,50.35) -- (418.1,62.99) -- (407.23,62.99) -- cycle ;
\draw    (412.67,73.83) ;
\draw [shift={(412.67,73.83)}, rotate = 0] [color={rgb, 255:red, 0; green, 0; blue, 0 }  ][fill={rgb, 255:red, 0; green, 0; blue, 0 }  ][line width=0.75]      (0, 0) circle [x radius= 3.35, y radius= 3.35]   ;
\draw    (399.83,84.67) -- (425.5,84.67) ;

\draw (209.75,38.21) node  []  {$n_{1} \ $};
\draw (173.75,40.71) node  []  {$s\ $};
\draw (405.75,38.21) node  []  {$n_{8} \ $};
\draw (375.75,38.21) node  []  {$n_{7} \ $};
\draw (345.75,38.21) node  []  {$n_{6} \ $};
\draw (323.75,38.21) node  []  {$n_{5} \ $};
\draw (293.75,38.21) node  []  {$n_{4} \ $};
\draw (263.75,38.21) node  []  {$n_{3} \ $};
\draw (238.75,38.21) node  []  {$n_{2} \ $};
\draw (428.75,38.21) node  []  {$n_{9} \ $};
\draw (455.75,38.21) node  []  {$n_{10} \ $};
\draw (444.75,54.71) node  [font=\small]  {Reservoir};
\draw (444.75,69.21) node  [font=\small]  {Junction};
\draw (444.75,83.71) node  [font=\small]  {Pipe};

\end{tikzpicture}
	}
	\vspace{-0.8cm}
	\caption{Exemplary Water Distribution Path considered for the test case}
	\vspace{-0.2cm}
	\label{fig:EPA_net}
\end{figure}
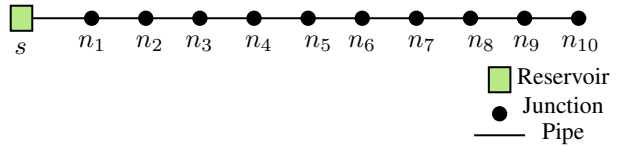

\begin{figure}[t!]
  \centering
    \includegraphics[width=\linewidth]{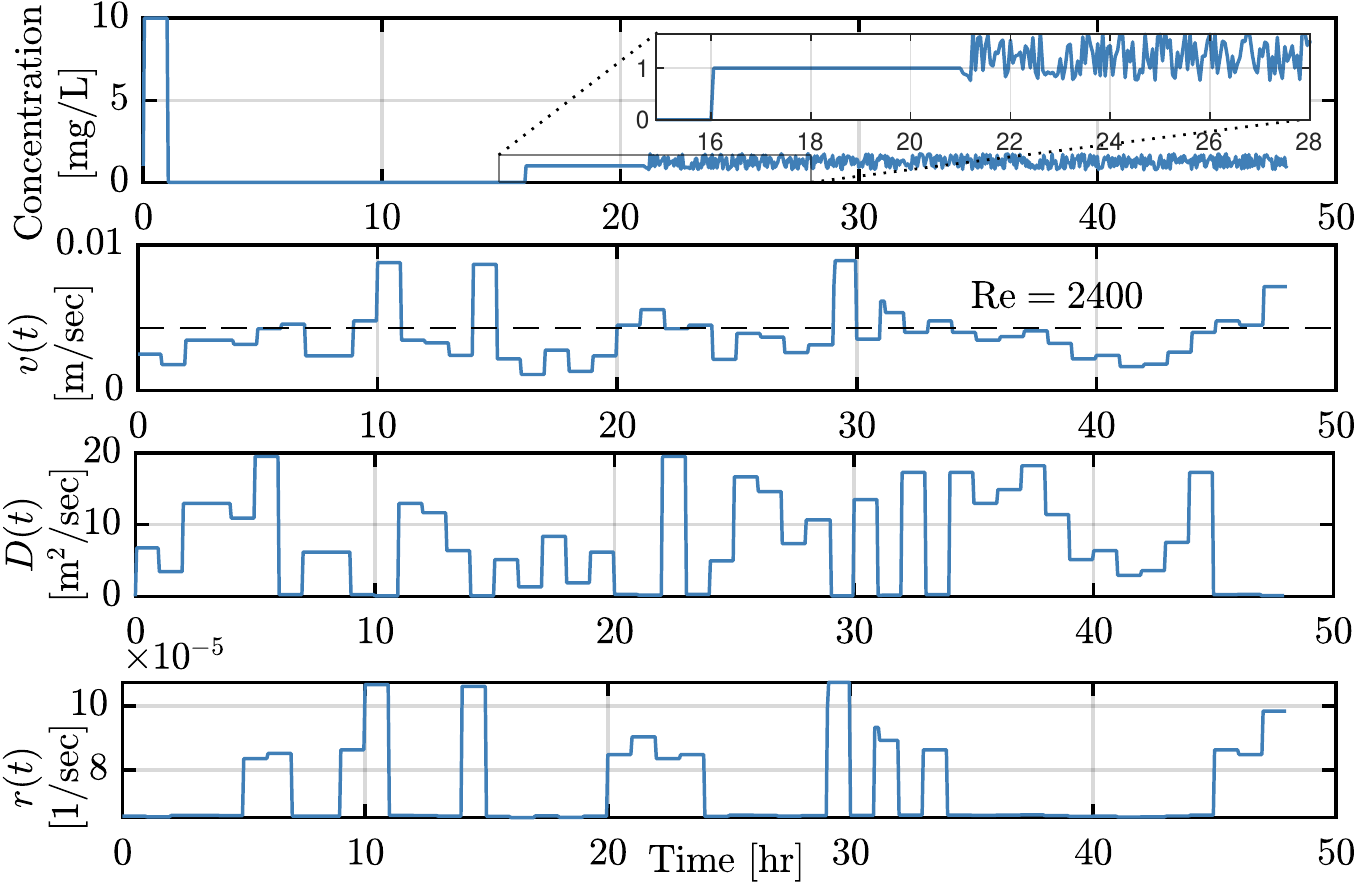}
    \vspace{-0.7cm}
  	\caption{Injected chlorine concentration profile (top) along with the ADR coefficients profiles used in the water quality simulation}
	\vspace{-0.6cm}
	\label{fig:Inprof}
\end{figure}

There are a number of aspects which could be considered for the modeling of the water quality such as the disinfectant concentration or biological activity of bacteria within the water. Here the popular disinfectant concentration aspect is considered. The disinfectant concentration within the water flowing in the pipe is modelled by a standard ADR PDE (\ref{eq:ADR_PDE}) with time varying coefficients. However,  the PDE coefficients in such case are interdependent. Namely, the axial diffusion coefficient $D(t)$ as well as the reaction rate $r(t)$ are functions of the flow regime, i.e. flow velocity $v(t)$. The exact dependency for the diffusion and reaction parameters $D(t),r(t)$ (and also effect of other variables as water temperature) is a separate research topic heavily relying on empirical studies \cite{DParamLee2004MASSDI,Rparam_kw,Dparam_Taylor_turb}. For the diffusion rate in laminar regime the relation proposed in \cite{DParamLee2004MASSDI} is used here. Moreover, for the turbulent regime the empirical formula achieved by the gene expression technique proposed in \cite{D_Gene_Exepr} is used due to its simplicity and acceptable accuracy. Hence, the total expression for the diffusion coefficient is as follows
\begin{equation*}
D(t)=\begin{cases}\frac{a^2v(t)^2}{48D_{0}}(1-\exp{-12.425\frac{\tau D_{0}}{a^2}}) & \Rey<2400 \\\frac{c_1(2v(t)+c_2)}{\Rey\,v(t)} & \Rey \geq 2400 
\end{cases}
\label{eq:DispWQ}
\end{equation*}
with the pipe radius $a$, Reynolds number $\Rey$, molecular diffusion coefficient $D_0$, the gene expression empirical formula constants $c_1,c_2$, and the Lagrangian time $\tau$, which is calculated in the simulation by averaging the velocity and dividing by the pipe total length $L$. The reaction coefficient $r(t)$ is decomposed into its two main contributors, the bulk reaction rate $k_b$ and the wall reaction $R_w(t)$. The classical dimensionless analysis relations of mass transfer \cite{Rossman1} are used to express the wall reaction rate in dependence of $v(t)$: 
\begin{subequations}
\begin{equation*}
r(t)=k_b+\underbrace{\frac{k_w k_f(t)}{a(k_w+k_f(t))}}_{R_w(t)},\, k_f=\Shear\frac{D_0}{2a},
\end{equation*}
\begin{equation*}
\Shear=\begin{cases}  3.65+\frac{0.0668(a/L)(\Rey\,\Schm)}{1+ [0.04(a/L)(\Rey\,\Schm)]^{2/3}} & \Rey <2400\\
0.023\,\Rey^{0.83}\Schm^{0.333}& \Rey\geq 2400 
\end{cases}
\end{equation*}
\end{subequations}
with the wall decay constant $k_w$, mass transfer coefficient $k_f$, Sherwood number $\Shear$, and Schmidt number $\Schm$. After constructing the model and its inputs dependencies, a ROM is constructed and used for the prediction of the disinfectant concentration within an exemplary water distribution path.
As one of the purposes of this contribution is to offer an enhancement to the existing water quality modeling and simulation tools, EPANET open source tool is used to construct the example. Moreover, the quality analysis produced by EPANET is compared to the one produced by the ROM and the high fidelity solution produced by MATLAB PDE solver highlighting the effect of considering the diffusion phenomena. The exemplary network is an adapted version of the one used in \cite{SHang_ADR} with consideration of a random demand pattern, which also ensures that both turbulent and laminar regimes are included. The network considered for the example is shown in Fig. \ref{fig:EPA_net}.
The network is composed of a source node, where chlorine concentration is injected, and 10 pipes with half meter diameter and length of 100 m (a total path of 1000 m). The chlorine concentration is initially set to zero then the profile shown in Fig. \ref{fig:Inprof} is injected for a 48 hour simulation time. The injected concentration profile includes a one hour pulse of 10 [mg/L] followed by a 16 hour zero injected concentration then finally a unit step superimposed by a random noise for the last 27 hours period. The case study includes periods for both laminar as well as turbulent flow regimes inducing the variability of the different regimes of the coefficients $D(t),r(t)$. Figure \ref{fig:Inprof} shows the profiles used for the 48 hour simulation, moreover the model parameters used for simulation are summarized in Table \ref{tb:Parmeter}.
\renewcommand{\arraystretch}{1}
\begin{table}[h!]
 \vspace{-0.4cm}
	\centering
	\caption{Parameter values used for the case study}
	\begin{tabular}{cccc}
		Parameter & Value & Unit\\\hline
		$a$&$0.5$&$\mathrm{m}$\\
		$D_0$&$1.2\times10^{-9}$& $\mathrm{m^2/sec}$\\
		$c_1$ & $4110$ & -\\
        $c_2$ & $0.062$ & -\\
        $k_b$ & $6.36\times 10^{-5}$ & $\mathrm{1/sec}$\\
        $k_w$ & $8.33\times 10^{-5}$ & $\mathrm{m/sec}$\\
        $\np$ & $500$ &-\\
        $n $ &$8$ &-\\
		\hline
	\end{tabular}	\label{tb:Parmeter}
	 \vspace{-0.4cm}
\end{table}

\section{Results}\label{sec:4}
The results of the two evaluation cases are summarized hereafter. Due to the limited place only few result plots for the first case will be presented.  

\subsection{Arbitrary Input Profiles Simulations}
\begin{figure}[h!]
  \centering
    \includegraphics[width=\linewidth]{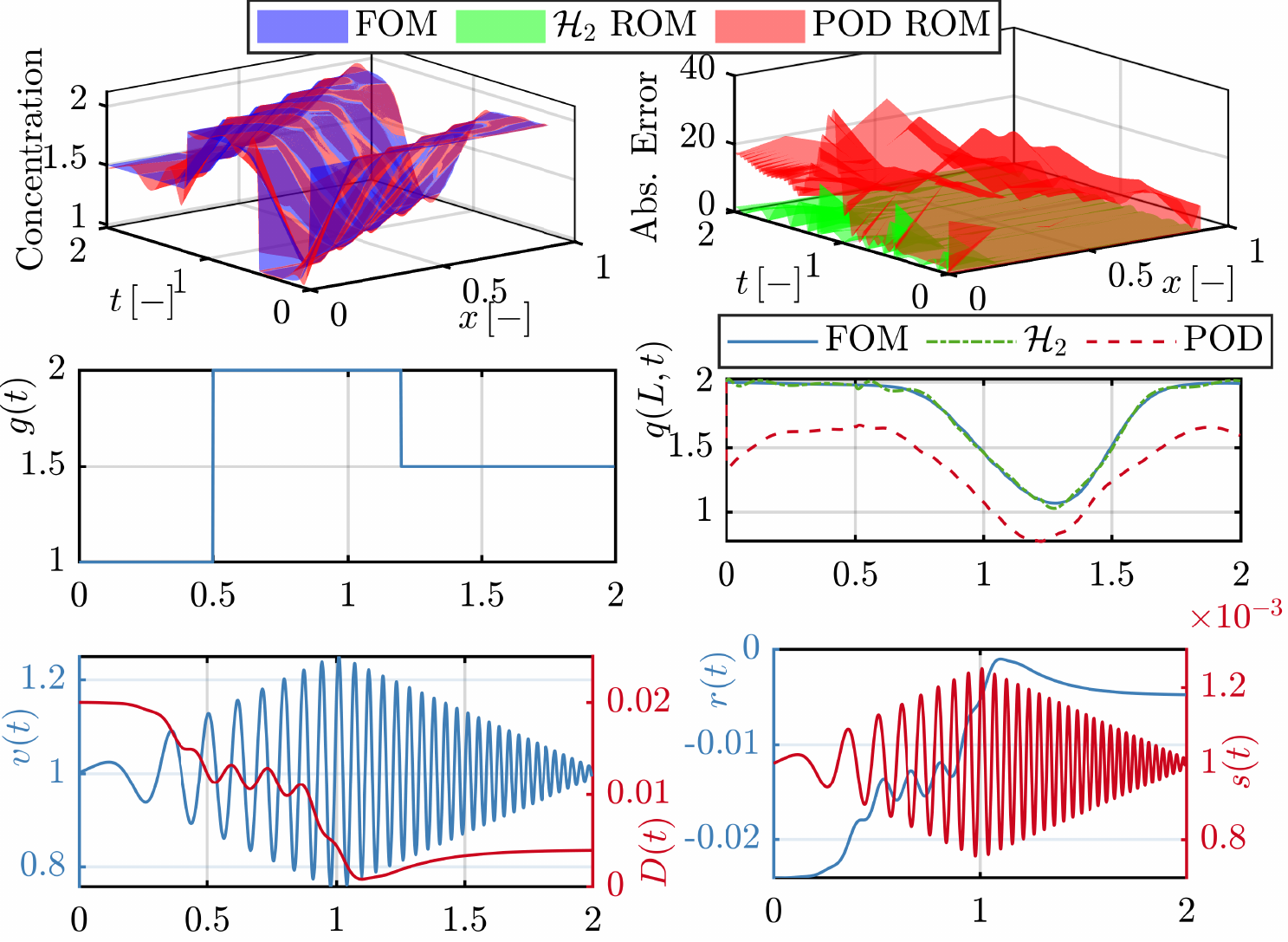}
    \vspace{-0.6cm}
  	\caption{The predicted concentration surface by the FOM (blue), $\mathcal{H}_2$ optimal ROM (green) and POD-Galerkin ROM (red) with the corresponding absolute error for a step change in injected concentration (middle left) and arbitrary changing profiles for $v(t),D(t),r(t),s(t)$ (bottom)}
	\vspace{-0.5cm}
	\label{fig:TRsurf}
\end{figure}

The results for one selected case are shown in Fig. \ref{fig:TRsurf}. Moreover, the deviation between the ROM and FOM prediction is quantified using Normalised Mean Square Error (NMSE) in percentage and summarized for all scenarios in Table \ref{tab:Results} (with the shown result in Fig. \ref{fig:TRsurf} highlighted in the table). As expected, the POD-Galerkin based ROM showed a large steady state error or unstable behavior for a large number of the chosen scenarios. This is as the POD-Galerkin based ROM performance is generally limited to the behavior introduced in the snapshots matrix. And for such rich system represented by the ADR system with the time variable coefficients it was not possible to capture the dependency (dynamic and static) of the varying coefficients fully in the snapshots matrix. On the other hand the ROM with the proposed $\mathcal{H}_2$ method could produce stable and accurate prediction for almost all cases for \peclet number up to $10^{5}$. Oscillatory behavior within the ROM behavior is observed for high \peclet number, i.e. $\Pecl>10^2$ which is expected due to the general limitation of the global bases methods in combination with extremely high \peclet\footnote{\peclet number is a non dimensional quantity equal to the ratio between rate of advection and rate of diffusion, hence, represents a quantification for the dominating flow phenomena.} number and Eulerian frame of reference (see \cite{Elkhashap.2021b}).
\newcommand{\vln}[1]{\multicolumn{1}{c|}{#1}}
\newcommand{\TabStack}[1]{{\multirow{2}{*}{\shortstack{#1}}}}
\definecolor{TabShade}{rgb}{0.00000,0.38039,0.39608}
\newcommand{\Hi}{\cellcolor{TabShade!20}}
\newcommand{\STAB}[1]{\begin{tabular}{@{}c@{}}#1\end{tabular}}
\begin{table}[h]
	\caption{Results of the two ROMs accuracy compared to the FOM for all simulations at different \peclet number}
	\vspace{-0.2cm}
	\setlength{\tabcolsep}{4pt} 
	\renewcommand{\arraystretch}{0.65} 
	\centering
	\begin{tabular}{cccccccc}
		\toprule
		& \multicolumn{1}{c}{\multirow{3}{*}{\begin{tabular}{@{}c@{}}$\Pecl$ \\ Range\end{tabular}}} & \multicolumn{3}{c}{\shortstack{$\mathcal{H}_2$ Norm Optimal ROM\\NMSE  $\UnitBrackets{\percent}$}} & \multicolumn{3}{c}{\shortstack{POD-Galerkin ROM\\ NMSE $\UnitBrackets{\percent}$}} \\ \cmidrule(r){3-5} \cmidrule(r){6-8}  
		& \multicolumn{1}{c}{}
		& Step  & Pulse  & Random  & Step  & Pulse  & Random  \\ \midrule
		\vln{\multirow{6}{*}{\STAB{\rotatebox[origin=c]{90}{$n=8$}}}} &
		\vln{$10^0$} &   $0.00009$&   $0.006$&   $0.0003$&   $207.6$&   $80.1$&   $30.2$\\	\vln{}&   
		\vln{$10^1$} &   $0.01726$&   $0.731$&   $0.0025$&   $167.9$&   $13.4$&   $5.70$\\	\vln{}&   
		\vln{$10^2$} &\Hi$0.24527$&   $9.355$&   $0.0219$&\Hi$75.39$&   $12.6$&   $5.90$\\	\vln{}&   
		\vln{$10^3$} &   $2.06860$&   $46.59$&   $0.4077$&  $64.57$&   $40.7$&   $6.42$\\	\vln{}&   
		\vln{$10^4$} &   $3.11280$&   $65.68$&   $0.6161$&   $63.84$&   $60.3$&   $6.71$\\	\vln{}&   
		\vln{$10^5$} &   $3.25190$&   $68.28$&   $0.8092$&   $63.79$&   $63.1$&   $6.77$\\  
		\bottomrule
	\end{tabular}\label{tab:Results}
\vspace{-0.5cm}
\end{table}

\subsection{Water Quality Modeling}
\begin{figure}[h!]
  \centering
    \includegraphics[width=\linewidth]{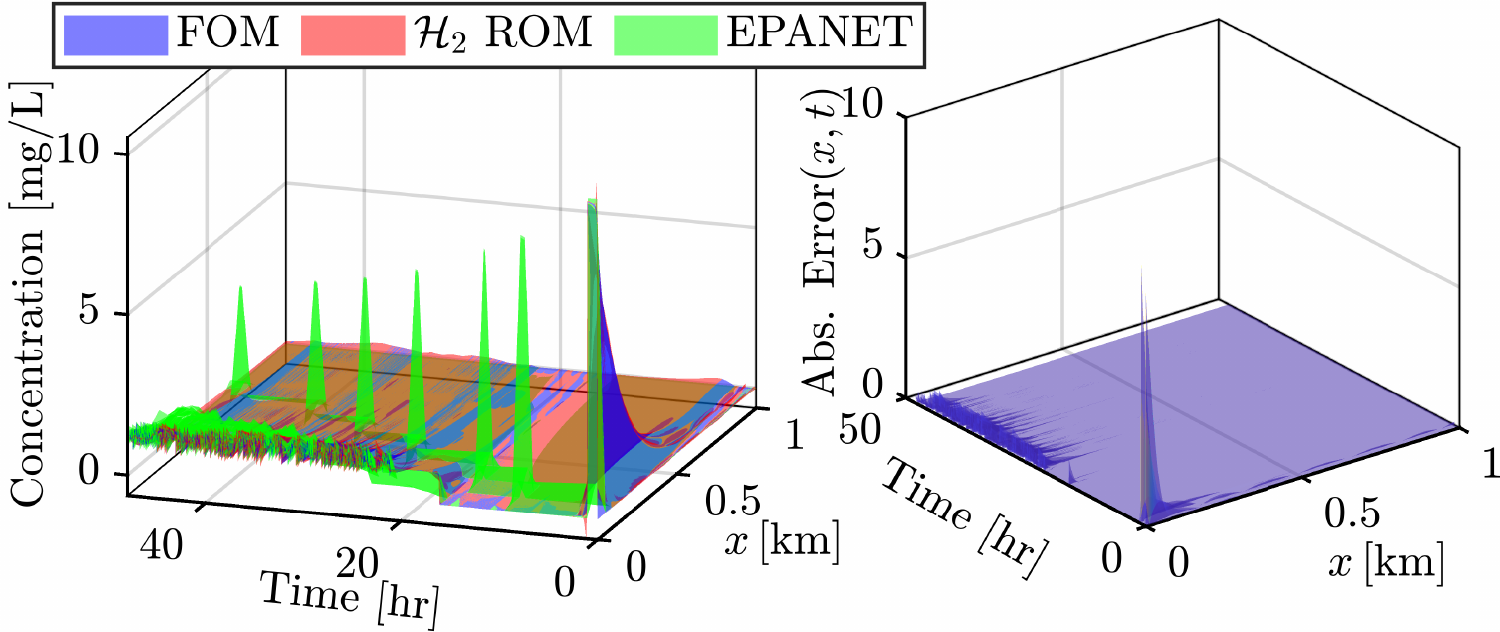}
    \vspace{-0.7cm}
  	\caption{The predicted concentration surface for the 48hr simulation time (left) with the ROM (red), high fidelity simulation (blue) and EPANET (green), absolute error between FOM and ROM (right)}
	\vspace{-0.2cm}
	\label{fig:WQsurf}
\end{figure}

\begin{figure}[h!]
  \centering
    \includegraphics[width=\linewidth]{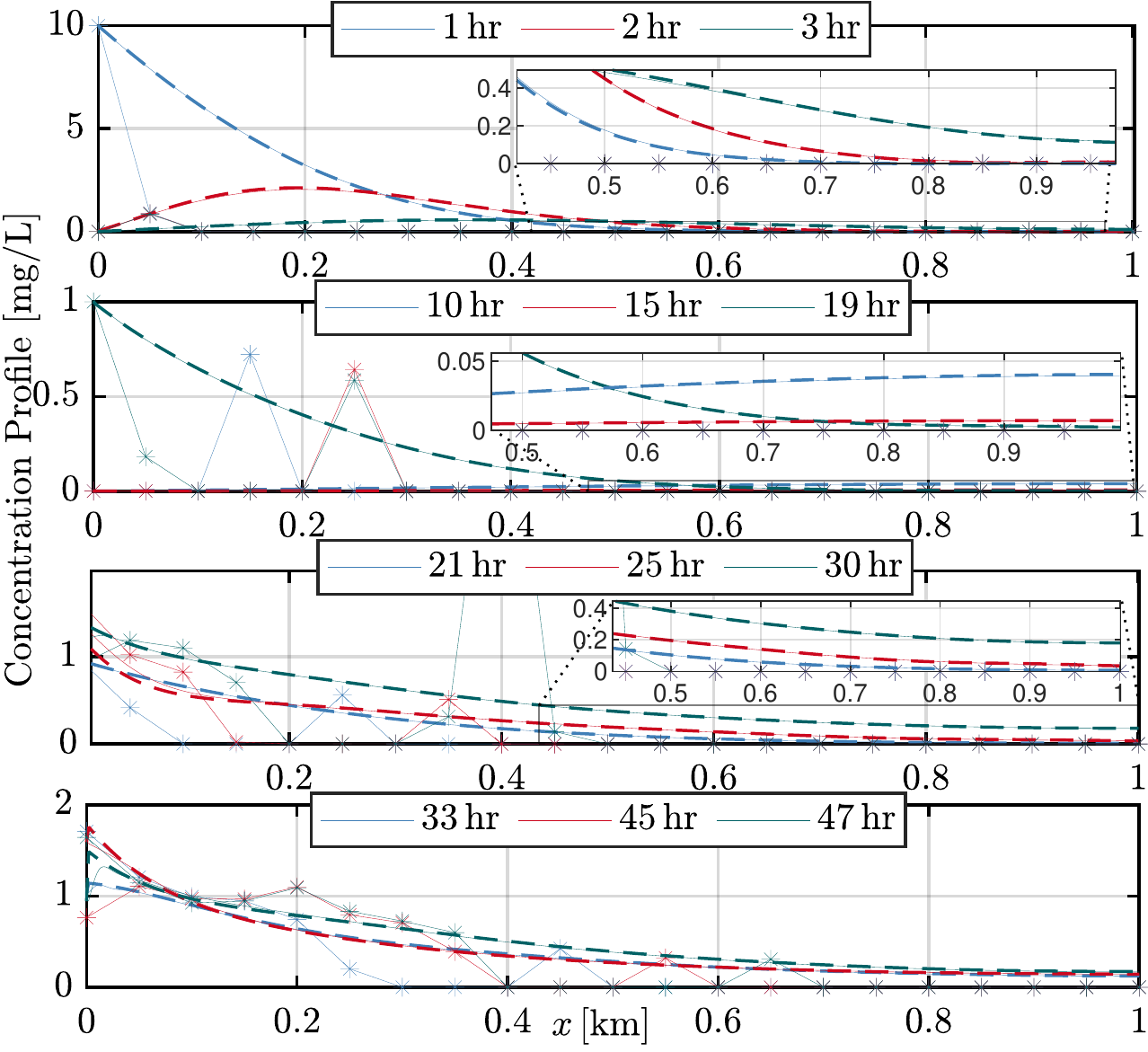}
    \vspace{-0.7cm}
  	\caption{The predicted concentration profiles with the ROM (dotted line), high fidelity simulation (solid line), and EPANET (asterisk Marker) at different time instants}
	\vspace{-0.8cm}
	\label{fig:Zplots}
\end{figure}

The FOM as well as ROM full spatial resolution prediction along with the error between both are shown in Fig. \ref{fig:WQsurf}. Moreover, the EPANET predicted quality plotted against both the FOM and ROM along the length of the water path (only nodes for EPANET) at certain time points are shown in Fig. \ref{fig:Zplots}.
It can be observed that the ROM maps the FOM prediction very well with minimal error mainly represented at the boundary (clf. Fig.\ref{fig:WQsurf}). Furthermore, the EPANET quality prediction can be seen as green spikes propagating through time and space in the left plot in Fig. \ref{fig:WQsurf}. This behavior is due to the combined effect of the sparse spatial resolution (10 nodes with 100 meter distance) and the neglection of the diffusive part of the flow. It can be also observed that the EPANET neglection of diffusion causes a significant deviation to the high fidelity simulation specially at periods of low \peclet number (dominating diffusion). For example underestimation of the concentration profile can be observed at the downstream nodes for several time instants, e.g. $t=2,3,19\, \mathrm{hr}$ (see Fig. \ref{fig:Zplots}), as the advective flow at these time instants has not yet propagated the concentration. However in reality as shown by the simulation including the diffusion phenomena, the concentration would have already diffused through these nodes at the specified times. On the other hand there are instants with overestimation of the concentration, e.g. $t=30\, \mathrm{hr}$ in Fig. \ref{fig:Zplots} (also observe peaks in Fig. \ref{fig:WQsurf}) as the neglected diffusion phenomena usually causes an evening out effect to sharp gradients contrary to pure advection wave propagation behavior. The deviation of the ROM from the high fidelity simulation is observed to be low for both regimes with very mild oscillations at certain time instants at abrupt changes near the left boundary (observe the error plot in Fig. \ref{fig:WQsurf}). 
The ROM NMSE is below $2.3\,\%$ indicating a high prediction accuracy. A computational time reduction from $3487.5409$ sec needed by the FOM to $0.1507$ sec needed by the ROM could be achieved (on a Windows 10 development computer, Intel(R) Core(TM) i7-7700HQ CPU@2.8GHZ, 8GB RAM). This indicates computational speedup reaching a factor of four orders of magnitude compared to the MATLAB's PDE toolbox. However, it is not fair to compare a ROM with predetermined fixed steps calculations with high fidelity scheme with step size adaption FEM tool. However, the ROM computational complexity indicates the real-time potential of the ROM. The computational time needed for the EPANET water quality analysis including the hydraulic analysis using the matlab interface is $2.576$ sec.

\section{CONCLUSIONS}
A model order reduction method for Advection Diffusion Reaction (ADR) PDEs with time varying coefficients is proposed. The method is based on formulating the PDE into a standard bilinear high dimensional ODE using semi-discretization and augmentation of the input vector. Then system theoretic projection based model order reduction method, namely $\mathcal{H}_2$ norm optimal method for bilinear system, is employed for reducing the high dimensional system. The method is evaluated for two simulation based test cases against a reference solution produced using MATLAB's Finite Element Method PDE solver. A real world water quality simulation scenario is constructed for the second evaluation scenario.The water quality prediction of the second scenario is also generated using EPANET tool and compared against the high spatial resolution prediction highlighting the advantage of considering the further effects of which the proposed method offers. Finally the ROM could achieve a significant speedup compared to the high fidelity simulation solving the 48 hour, 1000 m water quality prediction problem within $0.1507$ seconds with normalised mean square error below $2.3\,\%$. The shown results introduce the potential of combining selective model formulation with structure preserving model order reduction techniques in high spatial resolution real-time monitoring of water quality networks.  

\addtolength{\textheight}{-12cm}   









\bibliographystyle{IEEEtran}
\bibliography{root}

\begin{thebibliography}{10}
\providecommand{\url}[1]{#1}
\csname url@rmstyle\endcsname
\providecommand{\newblock}{\relax}
\providecommand{\bibinfo}[2]{#2}
\providecommand\BIBentrySTDinterwordspacing{\spaceskip=0pt\relax}
\providecommand\BIBentryALTinterwordstretchfactor{4}
\providecommand\BIBentryALTinterwordspacing{\spaceskip=\fontdimen2\font plus
\BIBentryALTinterwordstretchfactor\fontdimen3\font minus
  \fontdimen4\font\relax}
\providecommand\BIBforeignlanguage[2]{{%
\expandafter\ifx\csname l@#1\endcsname\relax
\typeout{** WARNING: IEEEtran.bst: No hyphenation pattern has been}%
\typeout{** loaded for the language `#1'. Using the pattern for}%
\typeout{** the default language instead.}%
\else
\language=\csname l@#1\endcsname
\fi
#2}}

\bibitem{Elkhashap.2019b}
A.~Elkhashap, R.~Meier, and D.~Abel, ``Modelling and control of a continuous
  vibrated fluidized bed dryer in pharmaceutical production,'' \emph{Die
  pharmazeutische Industrie : pharmind}, 2019.

\bibitem{Elkhashap.2021b}
A.~Elkhashap, D.~Rüschen, and D.~Abel, ``Distributed parameter modeling of
  fluid transmission lines,'' \emph{Journal of Process Control}, 2021, in
  press.

\bibitem{WQ_Epanet}
A.~Seyoum and T.~Tanyimboh, ``Integration of hydraulic and water quality
  modelling in distribution networks: Epanet-pmx,'' \emph{Water Resources
  Management}, vol.~31, 11 2017.

\bibitem{rossman1999epanet}
L.~A. Rossman, ``The epanet programmer's toolkit for analysis of water
  distribution systems,'' in \emph{WRPMD'99: Preparing for the 21st Century},
  1999, pp. 1--10.

\bibitem{RTP_Epanet_1}
P.~Ingeduld, \emph{Real-Time Forecasting with EPANET}, pp. 1--9.

\bibitem{SHang_ADR}
F.~Shang, H.~Woo, J.~B. Burkhardt, and R.~Murray, ``Lagrangian method to model
  advection-dispersion-reaction transport in drinking water pipe networks,''
  \emph{Journal of Water Resources Planning and Management}, vol. 147, no.~9,
  2021.

\bibitem{ABOKIFA2016107}
A.~A. Abokifa, Y.~J. Yang, C.~S. Lo, and P.~Biswas, ``Water quality modeling in
  the dead end sections of drinking water distribution networks,'' \emph{Water
  Research}, vol.~89, pp. 107--117, 2016.

\bibitem{matlabPDEtool}
\BIBentryALTinterwordspacing
MathWorks, ``Partial differential equation toolbox,'' 2020. [Online].
  Available: \url{https://www.mathworks.com/help/pde/index.html}
\BIBentrySTDinterwordspacing

\bibitem{WQModels4MPC}
S.~Wang, A.~F. Taha, and A.~A. Abokifa, ``How effective is model predictive
  control in real-time water quality regulation? state-space modeling and
  scalable control,'' \emph{Water Resources Research}, vol.~57, no.~5, 2021.

\bibitem{WQMOR}
S.~Wang, A.~F. Taha, A.~Chakrabarty, L.~Sela, and A.~Abokifa, ``Model order
  reduction for water quality dynamics,'' 2021.

\bibitem{MORbilinH2}
P.~Benner and T.~Breiten, ``Interpolation-based $\mathcal{H}_2$-model reduction
  of bilinear control systems,'' \emph{SIAM Journal on Matrix Analysis and
  Applications}, vol.~33, no.~3, pp. 859--885, 2012.

\bibitem{Elkhashap.2021a}
A.~Elkhashap and D.~Abel, ``Parametric model order reduction of variable
  parameter axial dispersion model,'' \emph{Conference on Control Technology
  and Applications (CCTA)}, 2021.

\bibitem{Elkhashap2022realtime}
A.~Elkhashap, D.~Rüschen, and D.~Abel, ``Towards real-time monitoring and
  control of water networks,'' in \emph{2022 American Control Conference
  (ACC)}, 2022.

\bibitem{redmann2021bilinear}
M.~Redmann, ``Bilinear systems -- a new link to $\mathcal{H}_2$-norms,
  relations to stochastic systems and further properties,'' 2021.

\bibitem{Danckwerts.1995}
P.~V. Danckwerts, ``Continuous flow systems. distribution of residence times,''
  \emph{Chemical engineering science}, vol.~50, no.~24, pp. 3857--3866, 1995.

\bibitem{Abgrall.2017}
R.~Abgrall, C.-W. Shu, and Q.~Du, Eds., \emph{Handbook of numerical methods for
  hyperbolic problems: Applied and modern issues}, ser. Handbook of numerical
  analysis.\hskip 1em plus 0.5em minus 0.4em\relax Amsterdam, The Netherlands
  and Kidlington, Oxford, United Kingdom: {North-Holland an imprint of
  Elsevier}, 2017, vol. volume 18.

\bibitem{MORbilinH2Zhang.2002}
L.~Zhang and J.~Lam, ``On h2 model reduction of bilinear systems,''
  \emph{Automatica}, vol.~38, no.~2, pp. 205--216, 2002.

\bibitem{Redmann2019TheML}
M.~Redmann, ``The missing link between the output and the $\mathcal{H}_2$-norm
  of bilinear systems,'' \emph{ArXiv}, vol. abs/1910.14427, 2019.

\bibitem{MORParamOverview}
P.~Benner, S.~Gugercin, and K.~Willcox, ``A survey of projection-based model
  reduction methods for parametric dynamical systems,'' \emph{SIAM Review},
  vol.~57, no.~4, pp. 483--531, 2015.

\bibitem{Andersson2019}
J.~A.~E. Andersson, J.~Gillis, G.~Horn, J.~B. Rawlings, and M.~Diehl,
  ``{CasADi} -- {A} software framework for nonlinear optimization and optimal
  control,'' \emph{Mathematical Programming Computation}, vol.~11, no.~1, pp.
  1--36, 2019.

\bibitem{DParamLee2004MASSDI}
Y.~Lee, ``Mass dispersion in intermittent laminar flow,'' Ph.D. dissertation,
  2004.

\bibitem{Rparam_kw}
G.~R. Munavalli, M.~S.~M. Kumar, and M.~A. Kulkarni, ``{Wall decay of chlorine
  in water distribution system},'' \emph{Journal of Water Supply: Research and
  Technology-Aqua}, vol.~58, no.~5, pp. 316--326, 08 2009.

\bibitem{Dparam_Taylor_turb}
G.~I. Taylor, ``The dispersion of matter in turbulent flow through a pipe,''
  \emph{Proceedings of the Royal Society of London. Series A. Mathematical and
  Physical Sciences}, vol. 223, no. 1155, pp. 446--468, 1954.

\bibitem{D_Gene_Exepr}
A.~M.~A. Sattar, ``Gene expression models for the prediction of longitudinal
  dispersion coefficients in transitional and turbulent pipe flow,''
  \emph{Journal of Pipeline Systems Engineering and Practice}, vol.~5, no.~1,
  2014.

\bibitem{Rossman1}
L.~A. Rossman, R.~M. Clark, and W.~M. Grayman, ``Modeling chlorine residuals in
  drinking water distribution systems,'' \emph{Journal of Environmental
  Engineering}, vol. 120, no.~4, pp. 803--820, 1994.

\end{thebibliography}

\end{document}